\documentstyle[12pt,aasms4]{article}

\def\la{\mathrel{\mathpalette\fun <}}
\def\ga{\mathrel{\mathpalette\fun >}}

\def\fun#1#2{\lower0.837ex\vbox{\baselineskip0ex\lineskip0.209ex
  \ialign{$\mathsurround=0ex#1\hfil##\hfil$\crcr#2\crcr\sim\crcr}}}

\def\msun{M_\odot}

\def\sles{\lower2pt\hbox{$\buildrel {\scriptstyle <}
   \over {\scriptstyle\sim}$}}
 
\def\sgreat{\lower2pt\hbox{$\buildrel {\scriptstyle >}
   \over {\scriptstyle\sim}$}}

\def\la{\mathrel{\mathpalette\fun <}}
\def\ga{\mathrel{\mathpalette\fun >}}

\begin{document}

 \title{ Ozone Depletion from Nearby Supernovae}

 \smallskip 

 \author{ Neil Gehrels }
 \affil{e-mail: gehrels@lheapop.gsfc.nasa.gov}
%\affil{Goddard Space Flight Center}
 \affil{NASA/GSFC/Laboratory for High Energy Astrophysics, 
 Code 661, Greenbelt, MD 20771}
 \authoraddr{NASA/GSFC/Laboratory for High Energy Astrophysics, 
 Code 661, Greenbelt, MD 20771}

\smallskip

 \author{ Claude M. Laird}
 \affil{e-mail:   claird@ku.edu}
 \affil{
  Department of Physics and Astronomy,
  University of Kansas,
  Lawrence, KS 66045}
 \authoraddr{
 Department of Physics and Astronomy,
  University of Kansas,
  Lawrence, KS 66045}

\smallskip

 \author{Charles H. Jackman}
 \affil{e-mail:  jackman@assess.gsfc.nasa.gov}
 \affil{NASA/GSFC/Laboratory for Atmospheres,
 Code 916, Greenbelt, MD 20771}
 \authoraddr{NASA/GSFC/Laboratory for Atmospheres,
 Code 916, Greenbelt, MD 20771}

\smallskip

 \author{ John K. Cannizzo\footnote{
also University of Maryland    Baltimore County}   }
 \affil{e-mail: cannizzo@stars.gsfc.nasa.gov}
 \affil{NASA/GSFC/Laboratory for High Energy Astrophysics, 
 Code 661, Greenbelt, MD 20771}
 \authoraddr{NASA/GSFC/Laboratory for High Energy Astrophysics, 
 Code 661, Greenbelt, MD 20771}

\smallskip

 \author{ Barbara J. Mattson\footnote{
also  L3 Com Analytics Corp.}  }
 \affil{e-mail: mattson@milkyway.gsfc.nasa.gov}
 \affil{NASA/GSFC/Laboratory for High Energy Astrophysics, 
 Code 661, Greenbelt, MD 20771}
 \authoraddr{NASA/GSFC/Laboratory for High Energy Astrophysics, 
 Code 661, Greenbelt, MD 20771}

\smallskip

 \author{ Wan Chen }
 \affil{e-mail:  Wan.W.Chen@mail.sprint.com }
 \affil{Sprint,  IP Design $-$ RAD,
  12502 Sunrise Valley Dr., 
  Reston, VA 20196} 

 \authoraddr{Sprint, IP Design $-$ RAD,
  12502 Sunrise Valley Dr., 
  Reston, VA 20196}

\centerline{to appear in the Astrophysical Journal 
 2003, March 10, vol. 585}

\received{ 2002 May 28}
\accepted{ 2002 November 12}

\begin{abstract}

Estimates made in the 1970's indicated
that a supernova occurring within tens of parsecs
of Earth could have significant effects on the ozone 
layer. Since that time improved tools for detailed
modeling of atmospheric chemistry have been 
developed to calculate ozone depletion,
and advances have been made also in theoretical
modeling of supernovae and of the resultant
gamma-ray spectra.
In addition, one now has better knowledge of
the occurrence rate of supernovae in the galaxy,
and of the spatial distribution of progenitors
to core-collapse supernovae.
We report here the results of two-dimensional
atmospheric model calculations that take as input
the spectral energy distribution of a supernova,
adopting various distances from Earth and various
latitude impact angles.
In separate simulations we calculate the ozone
depletion due to both gamma-rays and cosmic rays.
We find that for the combined ozone depletion
from these effects roughly to double the 
``biologically active'' UV flux
received
at the surface  of the Earth,
the supernova must occur at $\la 8$ pc. 
 Based on the latest data, the
           time-averaged
  galactic
  rate of core-collapse  supernovae  occurring within 8 pc
is $\sim1.5$ Gyr$^{-1}$.
   In comparing our calculated ozone depletions
 with those of previous studies,
   we find them to be  significantly  less
   severe  than found by Ruderman (1974),
and consistent with Whitten et al. (1976).
  In summary, given the amplitude of the effect, the
rate of nearby supernovae, and the
          $\sim0.5$ Gyr time scale for multicellular
  organisms on Earth,  this particular pathway
     for mass extinctions may be 
less important than previously thought.

\end{abstract}

\medskip
\medskip

{\it  Subject headings:}
molecular processes; Earth; stars: supernovae: general;
  supernovae: SN1978A; 
   ISM: cosmic rays

\section{ Introduction }

Ruderman (1974) suggested and
      was the first to study the reduction
of stratospheric $O_3$ due to enhanced
levels of nitrogen oxides caused by incident
radiation from a supernova (SN).
With rough calculations he found that a nearby ($<17$ pc)
SN would cause a reduction of  $O_3$ by $\sim80$\%
for $>2$ yr from the gamma radiation,
and a $40-90$\% reduction in  $O_3$  lasting
hundreds of years from cosmic rays (Ruderman 1974, Laster 1968).
 Reid et al. (1978) reached similar conclusions,
while Whitten et al. (1976)
found a much smaller effect.
 Ellis \& Schramm (1995) present a study similar
to Ruderman's in which simple analytical scalings were
utilized. Crutzen \& Br\"uhl (1996) also examine the
problem using a time dependent two-dimensional (2D) model,
  considering only the enhanced cosmic ray flux.
Except for  Crutzen \& Br\"uhl,
   these studies used simple theoretical models for
the SN energy and had, at best, one-dimensional (1D)
photochemistry models. 
Advances in computing power and atmospheric models
now make a more detailed analysis possible.
Also, SN1987A observations and recent state-of-the-art calculations
provide an estimate for the total gamma-ray input from a core-collapse
SN.

\section {Methodology}

We model  the effects of a core-collapse SN on the Earth's
ozone by inputting the expected cosmic and gamma irradiation
into an atmospheric model.
We consider: (1) the relatively short-lived ($\sim100$ d) gamma
rays from the initial blast, and
(2) the longer lived ($\ga 10$ y) cosmic rays accelerated
in the SN blast wave.
%   In our calculations, run times of 20 yr for both (separate)
%sets of calculations were chosen so as to ensure achieving
%a steady state. The actual duration of the cosmic rays from a 
%SN is uncertain, and $\sim10-20$ yr is toward the lower range
%of the values which have appeared in the literature.
We examine the effects of SN distance and impact angle
latitude, and determine the ozone
depletion averaged latitudinally and globally.

\subsection {Atmospheric Model}
We use the NASA/Goddard Space Flight Center (GSFC)
two-dimensional photochemical transport model, whose two dimensions
are latitude and altitude.  The latitude range is from the North
pole to the South pole and from the ground up to about 116 km.
The GSFC model has 18 latitude bands with a
latitude grid spacing of 10 degrees.  The altitude range includes 
58 evenly  spaced logarithmic pressure levels (approximately 2 km 
vertical grid point separation).

This model, including information about the chemistry and
computational approach, was originally described in Douglass et al.
(1989) and had an altitude range up to 60 km that included the 
troposphere, stratosphere, and lower mesosphere.  The altitude 
range was extended through the mesosphere
with the inclusion of several additional photochemical reactions,
which were discussed in Jackman et al. (1990).  Heterogeneous
processes occurring on the stratospheric sulfate aerosol layer
and on polar stratospheric clouds are important in the
stratosphere and are included in the manner described in
Considine et al. (1994).

The model uses a look-up table for computation of the photolytic
source term, which is employed in calculations of photodissociation rates
of atmospheric constituents from sunlight (see Jackman et al.,
1996).  Reaction rates and photolysis cross sections in the model
are consistent with the Jet Propulsion Laboratory recommendations
(DeMore et al., 1997).

The model includes both winds and small scale mixing processes in 
the manner described in Fleming et al. (1999), which is
briefly reiterated here.  The
meridional (north-south) and vertical (up-down) winds are
climatological in nature.  A 17-year average (1979-1995) of 
temperature data from the National Centers for Environmental 
Prediction (NCEP) and heating rates from climatological distributions
of temperature, ozone, and water vapor are the primary
ingredients in deriving the winds.  The small scale horizontal
mixing (eddy diffusion) coefficients are obtained self-consistently
from the winds and the 17-year NCEP analyses for planetary wave
climatology.  The small scale vertical mixing (eddy diffusion)
coefficients are computed from the mechanical forcing of gravity
waves in the mesosphere and upper stratosphere and from the
vertical temperature gradient in the troposphere and lower
stratosphere.

The model includes a new numerical advection
scheme detailed in Fleming et al. (1999), which is mass conserving
and utilizes an upstream piecewise parabolic method of solution.
A time step of 12 hr is employed for the advection of constituents 
and a time step of 1 d is used for the computations of 
changes in the constituents resulting from photochemical reactions. 

Recent work with the GSFC 2D model has focused on the influence
of solar proton events (SPEs) on atmospheric constituents.
Vitt \& Jackman (1996) modified the model to incorporate energy 
deposition by solar protons following Armstrong et al. (1989).

The impact of galactic cosmic rays (GCRs) on the atmosphere
was introduced into the model following Nicolet (1975).  The
Nicolet (1975) study relied on measurements of Neher (1961, 1967,
1971), who made a number of balloon flights which carried instruments 
to measure ionization rates at various latitudes and pressure.  Nicolet 
used  these data to compile the ionization rate
(in units of cm$^{-3}$ s$^{-1}$) as a function
of geomagnetic latitude, altitude, and phase (maximum or minimum)
of the solar cycle.  These data provide the most reliable estimates
of atmospheric ionization rates.

\subsection {Inclusion of SN-produced Gamma- and Cosmic Rays}
 The observed gamma-ray
% photon
     spectrum
  for
SN1987A is

\begin{equation}
 {dN\over dE } = 1.7\times 10^{-3} \ \left(E \over {1 \ {\rm MeV}}  \right)^{-1.2} 
 \ \ {\rm cm}^{-1} \ {\rm s}^{-1} \ {\rm MeV}^{-1}
\end{equation}

\noindent (Gehrels et al. 1988)
between 0.02 and 2 MeV, lasting 500 d at 55 kpc, 
for a total energy $9.0\times 10^{46}$ erg.
For ease of modeling 
we set the incident monoenergetic gamma-ray photon
flux  $N_i^0$  by binning this differential flux into
66 evenly spaced logarithmic intervals from 0.001 to 10 MeV,
for a net energy input of $3.3\times 10^{47}$ erg.
For a given distance  $D_{\rm SN}$
  we scale the empirically observed SN1987A spectrum
   by  $(5.5\times10^4/D_{\rm SN})^2$,
with $D_{\rm SN}$ in pc.
   In addition, in our final analysis we rescale our results
to a total gamma-ray energy of  $1.8\times 10^{47}$ erg
because of the following:
%for reasons discussed in the next paragraph.
%
%
SN1987A was unusual in that its progenitor
was a blue supergiant rather than the more typical
red supergiant.
A recent three-dimensional (3D) smooth particle hydrodynamics (SPH) 
calculation\footnote{Aimee Hungerford kindly 
   provided the results of her SN calculations
  in advance of publication.
  Her initial model, s15s7b from Weaver \& Woosley (1993),
was taken 100 s after bounce and  mapped
into a 3D SPH code.
  To calculate the gamma-ray spectra and total energies,
she mapped time slices from the SN calculation
    into a 3D Monte Carlo gamma-ray transport code.}
of a SN with a $15\msun$ red supergiant progenitor
gives $\sim[1.8\pm0.7]\times 10^{47}$ erg
for models which are initially spherically symmetric.
  In this model, the gamma-ray luminosity peaks
at $t\sim340$ d
and is within a factor of 10 of the peak
for $\sim500$ d.
 Since the energies of the gamma-ray photons are
so much greater than those involved in the atmospheric
chemistry reactions, the total input energy is
% in some sense
more relevant than the detailed SN spectrum
and we therefore have done all scalings
 using integrated energy inputs.

Treating the energy bins as monoenergetic beams
of geometric mean energy $<E_i>$
and integrating over a range of energies
(from $i=1$ to 66), the incident photon flux is given by
\begin{equation}
 N_i^0 = 8.5\times 10^{-3} \left[E_i^{-0.2} - E_{i+1}^{-0.2}\right] 
    \ {\rm cm}^{-2} \ {\rm s}^{-1} 
\end{equation}
\noindent 
where the total incident energy flux in the monoenergetic
  beam at the top of the atmosphere is $F_i^0 = N_i^0 <E_i>$.
We attenuate the gamma-ray  photon flux with altitude via an exponential
decay law, with the frequency-dependent absorption
coefficient taken from a look-up table (Plechaty et al. 1981).
 The beam is propagated vertically through a standard
atmosphere, which is adjusted later in the photochemistry
model for the appropriate latitude and time of year.
  The photon flux remaining in a monoenergetic
  beam is given by  $N_{i,j} =N_i^0 e^{-\mu_i x_j}$,
where $x_j$ is the column density (in g cm$^{-2}$) measured
from the top of the atmosphere and $\mu_i$ is the mass
attenuation or absorption coefficient for the $i$th energy bin
(i.e., associated with energy $<E_i>$).
 The photon flux, $\Delta N_{i,j}$ (in photons cm$^{-2}$ s$^{-1}$),
deposited in the $j$th
layer with energy $<E_i>$, is the
difference between $ N_{i,j-1}$ and  $ N_{i,j}$.
The energy flux deposited in the $j$th layer
by photons of energy  $<E_i>$ is $F_{i,j} = \Delta N_{i,j} <E_i>$
(in MeV cm$^{-2}$ s$^{-1}$).
  The total ionization rate is the sum over all energies

\begin{equation}
q_{{\rm tot,} \ j}  = {1\over {35 \ {\rm eV}}} \sum_{i=1}^{66} { {F_{i,j}}\over
  {\Delta Z_j}}, 
\end{equation}

\noindent 
   where 35 eV is the energy required to produce an ion pair (Porter et al.
 1976) and $\Delta Z_j$ is the thickness of the $j$th slab.

     The energy deposition versus altitude
calculated in this way is not dramatically
different from that obtained using
    a detailed radiative transfer
model.\footnote{
We thank David Smith, John Scalo, and Craig Wheeler
for
kindly sharing results from their Monte Carlo atmospheric
radiative transfer calculation.
The robustness of adopting simple energy-dependent
attenuation
coefficients
   in terms of the energy deposition versus altitude
was first shown by Chapman (1931).}
  Significant energy deposition into the atmosphere below some
altitude $h_{\rm crit} \sim16-20$ km, depending on latitude, will 
tend to create ozone rather than to destroy it, therefore
one basic check on our assumption of adopting
frequency-dependent attenuation coefficients
is to compare our fractional energy deposition versus altitude
     against the
  radiative transfer calculations of Smith et al.
(private communication). We
estimate the altitude-dependent
   energy deposition in our model by taking the
Earth's atmosphere to be exponential 
 $\rho = \rho_0 \exp(-h/H_{\rho})$,
where the sea level density $\rho_0 = 1.3\times 10^{-3}$ g cm$^{-3}$
and $H_{\rho} = 8$ km. (This gives an integrated
column density $\Sigma = \rho_0 H_{\rho} = 1040$ g cm$^{-2}$.)
Using the frequency-dependent
absorption coefficients from Plechaty et al. (1981)
shows that, for photon energies between 0.25 MeV
and 2. MeV, for example, the peak energy deposition
varies between about 30 and 40 km, well above  $h_{\rm crit}$.
The Smith et al. energy deposition at 1 MeV peaks at
$\sim 32$ km.  Finally, a detailed comparison of the
ratio of fractional energy deposition versus altitude
at 1 MeV
   between our model and Smith et al. shows the ratio to be
within $\sim0.2$ dex of unity down to $h\simeq 30$ km,
and above unity at smaller altitudes where the relative
energy deposition drops asymptotically.
In the model of Smith et al.,
  the altitude of maximum energy deposition decreases
from $\sim34$ km to $\sim28$ km
   as the photon energy varies between 0.25 MeV and 2 MeV,
   while  over this same energy range 
  our altitude of maximum energy deposition decreases
 from $\sim38$ km to $\sim30$ km.
  Thus there does not appear to be a significant quantitative
difference in the relative energy depositions
% below $h_{\rm crit}$
 between the two methods in the region where the dominant deposition
occurs $-$ a region which fortuitously lies above $h_{\rm crit}$
in both sets of calculations.

We estimate the SN cosmic ray inputs from the galactic
cosmic ray (GCR)
ionization rate profiles
at various latitudes
and solar cycle phase (Vitt \& Jackman 1996, Nicolet 1975).
The mean rate of nitrogen atom ($N$) production
at different altitudes and latitudes
is computed by  multiplying the empirically computed
rate of ionization (from Nicolet 1975) by 1.25 (Jackman et al. 1990).
The $N$ production enhances nitrogen oxide amounts as well as total
odd nitrogen, $NO_y$ (e.g., $N$, $NO$, $NO_2$, $NO_3$, $N_2O_5$,
$HNO_3$, $HO_2NO_2$, $ClONO_2$, and $BrONO_2$).
 The local GCR energy density $\sim1$ eV cm$^{-3}$
is dominated by protons with a peak at $\sim0.5$ GeV (Webber 1998), 
 implying a flux $\sim7$ cm$^{-2}$ s$^{-1}$.
Each incident cosmic ray produces $\sim10^7$ ionizing secondary
particles. The GCR  ionization rate inputs (from Nicolet 1975)
were multiplied by 100 to simulate the charged particle flux
from a SN at 10 pc, and scaled by $(10 \ {\rm pc}/D_{\rm SN})^2$
for other distances. This multiplicative scaling factor
was chosen simply and solely 
             to give a total SN energy in cosmic rays
   consistent with what is currently
thought to be a representative value.
For $D_{\rm SN} = 10$ pc, the locally observed GCR flux $\sim0.05$
erg cm$^{-2}$ s$^{-1}$ becomes $\sim5$ erg cm$^{-2}$ s$^{-1}$,
yielding a total energy  $\sim4\times 10^{49}$ erg over a 20 y run.
 The corresponding fluence is  $\sim3\times 10^9$  erg cm$^{-2}$.

The vertical ionization rate profiles
are then mapped onto a 2D circular grid, representing a zonally
averaged spherical Earth.
 First a series of ionization rate profiles as
functions of atmospheric slab number and latitude are calculated
for a flat Earth for incidence angles of 5$^\circ$, 15$^\circ$,...,
85$^\circ$ from the zenith, corresponding to 10$^\circ$ wide latitude
bands centered on 85$^\circ$, 75$^\circ$,..., 5$^\circ$, respectively,
for the illuminated hemisphere.
These profiles are interpolated to produce vertical average ionization
profiles as functions of altitude and latitude.
The zenith frame daily ionization profiles are mapped into 
360 intervals of 1$^\circ$ longitude and 18 intervals of 10$^\circ$ latitude.
  Zonally averaged daily  ionization rate profiles are produced
by averaging over all longitudes for each  10$^\circ$ 
latitude band, and then input into the GSFC 2D model.

\subsection {Atmospheric Model Simulations}

Three types of calculations of 20 y
duration were carried out: (1) a ``base'' simulation with no cosmic or gamma
rays, (2) ``perturbed'' simulations with enhanced cosmic ray levels, and
(3) ``perturbed'' simulations with enhanced gamma-rays.
The   ``perturbed'' simulations were compared to the ``base''
simulation to assess the atmospheric changes.
For the cosmic ray trials, the charged particle fluxes
are held constant for 20 y,
and the results are then checked to ensure a steady state
is achieved.
  For the gamma-ray trials, the irradiation is activated
on day 60 (March 1) and maintained at a constant level
for 300 d (the SN duration), after which it is attenuated
abruptly while the model is run for another 19 y,
allowing the perturbed atmosphere time to relax
to pre-SN conditions.

\section {Results}

Atmospheric ozone is easily produced through dissociation of
molecular oxygen.  The reaction

\hspace{0.5in}$O_2$ + h$\nu$($<$242 nm) $\rightarrow$ $O$ + $O$ \hspace{0.5in}\hfil  (A)

%followed by the three-body reactions
 followed by the three-body reaction

\hspace{0.5in}$O$ + $O_2$ + $M$ $\rightarrow$ $O_3$ + $M$     \hspace{0.5in}\hfil  (B)

%\begin{flushleft}
%\hspace{0.3in}and $O$ + $O_2$ + $M$ $\rightarrow$ $O_3$ + $M$ \hspace{0.5in}\hfil
%\end{flushleft}

\noindent
illustrates this production process. 
 Reaction (B) as written is schematically representative
  of the fact that
molecular oxygen ($O_2$)
is photodissociated into two atoms of oxygen ($O$), and each of these
attaches with an $O_2$ to form ozone ($O_3$),
    thereby producing  two ozone molecules.
 The $M$ in reaction (B) represents a
third body, which is usually either $N_2$ (78\% of the atmosphere)
or $O_2$ (21\% of the atmosphere).

Ozone is lost through a number
of catalytic reactions involving several ``families'' or groups
of constituents, such as $NO_y$, $HO_x$ (e.g., $H$, $OH$, 
and $HO_2$), $Cl_y$ (chlorine-containing inorganic molecules), 
and $Br_y$ (bromine-containing inorganic molecules).  These
constituents have both natural and human-made sources, which
are explained in several books (e.g., Dessler 2000).

Odd nitrogen, $NO_y$, is primarily created through natural
processes.  The major source of $NO_y$ is the oxidation of
biologically-produced $N_2O$ in the stratosphere 
(e.g., Vitt \& Jackman 1996). 
The $NO_y$ constituents can destroy ozone through
the catalytic reaction cycle

\hspace{0.5in}$NO$ + $O_3$ $\rightarrow$ $NO_2$ + $O_2$ \hspace{0.5in}\hfil  (C)

\hspace{0.5in}$NO_2$ + $O$ $\rightarrow$ $NO$ + $O_2$ \hspace{0.5in}\hfil  (D)

\hspace{0.3in}Net: $O_3$ + $O$ $\rightarrow$ $O_2$ + $O_2$ \hspace{0.5in}\hfil (E)

Note that $NO$ is not consumed through reactions (C) and (D) and
the net result is combining an ozone molecule and an atom of
oxygen to form two molecules of oxygen.  This cycle involving
these two reactions can proceed several hundred 
times before either $NO$ or $NO_2$ reacts with another
atmospheric constituent.

All the ``perturbed'' simulations include excess $NO_y$ from either 
cosmic or gamma-rays.  We describe the atmospheric effect from
these impacts, primarily focusing on the $NO_y$ and $O_3$ induced
variations.

Five gamma-ray simulations were performed
for SN impact angle latitude (i.e., latitude
for which the SN is at zenith)
$i_{\rm SN}=-90^{\circ}, \ -45^{\circ},  \ 0^{\circ},
       \ +45^{\circ},$ and $+90^{\circ}$,
taking $D_{\rm SN} =10$ pc.
Three more simulations were carried out for $i_{\rm SN}=0^{\circ}$
and $D_{\rm SN} =20$, 50, and 100 pc.
We find that after 300 d of simulated gamma radiation input,
the calculated
changes in $NO_y$ and $O_3$
  column
density  depend significantly on $i_{\rm SN}$.
  The largest increases
  in  $NO_y$  for $D_{\rm SN} =10$ pc are
for $i_{\rm SN}=\pm90^{\circ}$, 
and exceed 1800\% over the poles.
  The smallest increases occur for  $i_{\rm SN}=0^{\circ}$,
where the maximum increase is $<900$\% by the end of the run.
Ozone depletion accompanying the increase in
    $NO_y$ is greatest for $i_{\rm SN}=\pm90^{\circ}$,
with the zone of maximum decrease ($\sim60$\%)
at 60-70$^{\circ}$ S
latitude for $i_{\rm SN}= -90^{\circ}$.
The smallest decrease occurs for $i_{\rm SN}= 0^{\circ}$,
with the region of maximum decrease (about 40\%)
at  60-90$^{\circ}$ S latitude (Figures 1 and 2).
The maximum annual global average decrease for $D_{\rm SN} =10$ pc
(Figure 3) is 27\% in Year 2 for $i_{\rm SN}= 0^{\circ}$;
the minimum is 18\% for $i_{\rm SN}= -90^{\circ}$.
Simulations for 10, 20, 50, and 100 pc
show a $\sim D_{\rm SN}^{-n}$
trend in ozone depletion, where $\sim1.3 < n < \sim1.9$ (Figure 4).
In all cases, the global average ozone depletion from enhanced
gamma-rays decreases to a few percent by the sixth or seventh year.

        Figure  4 shows an increasing deviation
from a $D_{\rm SN}^{-2}$
law
  for the ozone depletion as $D_{\rm SN}$
decreases. In other words,
  although the fluence or net energy input from
gamma-rays  varies (by assumption) as $D_{\rm SN}^{-2}$,
  the ozone depletion begins to saturate
for small $D_{\rm SN}$.
 Three processes lead to this deviation: 

\hspace{0.5in}1) High production 
levels of $NO_y$ result in large self-destruction of $NO_y$
and will somewhat limit the modeled concentrations of $NO_y$
(Crutzen \& Br\"uhl 1996). 
The $N$ atoms produced via dissociation of $N_2$ primarily
react to form $NO$ via

\hspace{0.5in}$N$ + $O_2$ $\rightarrow$ $NO$ + $O$ \hspace{0.5in}\hfil  (F)

\begin{flushleft}
however, at higher levels of $NO$ the following reaction is
also important
\end{flushleft}

\hspace{0.5in}$N$ + $NO$ $\rightarrow$ $N_2$ + $O$ \hspace{0.5in}\hfil  (G)

\begin{flushleft}
This reaction brings nitrogen out of the odd nitrogen ($NO_y$) family 
back to even nitrogen, $N_2$.  
\end{flushleft}

\hspace{0.5in}2) Ozone can be produced (rather than destroyed)
 by enhancements
of $NO_y$ 
      in the troposphere and in the lowest part of the tropical
stratosphere (Crutzen \& Br\"uhl 1996).  The reaction 

\hspace{0.5in}$NO$ + $HO_2$ $\rightarrow$ $NO_2$ + $OH$ \hspace{0.5in}\hfil  (H)

\begin{flushleft}
repartitions constituents within the $NO_y$ and $HO_x$ families.
$NO_2$ is rather easily dissociated through
\end{flushleft}

\hspace{0.5in}$NO_2$ + h$\nu$($<$423 nm) $\rightarrow$ $NO$ + $O$ \hspace{0.5in}\hfil (I)

\begin{flushleft}
and $O_3$ is then formed via the three-body reaction (B).
\end{flushleft}

\hspace{0.5in}3) Finally, the enhanced $NO_y$
begins to interfere with other families (e.g., chlorine-, bromine-, and
hydrogen-containing constituents) that destroy ozone, thus reducing
the resultant ozone destruction from those families (e.g., this 
mechanism for production of $NO_y$ due to extremely large solar proton
events was discussed in Jackman et al. 2000).  Reactions such as

\hspace{0.5in}$ClO$ + $NO_2$ + $M$ $\rightarrow$ $ClONO_2$ + $M$ \hspace{0.5in}\hfil (J)

\begin{flushleft}
become important.  
\end{flushleft}

For cosmic ray simulations, we find that after achieving
the perturbed steady state, $NO_y$ column densities increase
at all  latitudes and for all times of the year.
For $D_{\rm SN} =10$ pc, the largest increases in column $NO_y$
(650\%) occur over the North Pole in late summer, while the
smallest increases (200\%) occur over the South Pole in winter,
with an average global increase of about 350\%.
  Corresponding column $O_3$ levels decrease globally and in all
seasons by as much as 40\% over the North Pole, in late summer
and by as little as 5\% over the equator during May-August for an
average global decrease of 22\%. 
  Our global ozone reductions are less than those of Ruderman (1974)
and Reid et al. (1978),  but consistent with Whitten et al. (1976).

\section {Uncertainties }

%There are many uncertainties in these computations, both in
%the input and the output of model results.  We discuss some
%of these uncertainties below.

The GSFC 2D model has
been developed to represent the ``present-day'' atmosphere.
Winds and small-scale mixing processes rely on an average of
meteorological field measurements over the 1979-1995 time
period and contemporary constituent measurements (see section 2.1).
%
%
%
%Additional sentences: "
The Earth's atmosphere has changed dramatically
since life first evolved about 3.5 Gyr  ago.  It has only been
since the Cambrian period began ($\sim 0.55$ Gyr ago) that molecular oxygen,
and  hence  ozone,
      have become significant components of the Earth's 
atmosphere.  Our computations are, therefore, not applicable to any
period before about  0.55 Gyr ago.

If the base state ozone in the past atmosphere, which would
have been perturbed by a SN,  were radically different from
 that indicated by
recent measured levels of ozone, then the modeled transport
field could be fairly different from the actual
past atmosphere.  Fleming et al. (2001) discussed differences
in the total ozone perturbation response from a stratospheric
aircraft perturbation.  The studied aircraft perturbations
were somewhat similar to a SN perturbation since the 
major atmospheric impact was
in the lower stratosphere.  Fleming et al. (2001) show about
a factor of two variation in the global total ozone perturbation
response.

Another possible concern relating to the atmospheric transport is
 the input energy from cosmic and gamma-rays.  This excess
energy or ``atmospheric heating'' can be quantified and compared
with background levels of heating:  At an altitude of about
15 km or a pressure of about 100 hPa, the background heating
rate is $\sim 0.1- 0.2$ K d$^{-1}$
(Joan Rosenfield, 1994
personal communication).
%(Rosenfield et al. 1994)
     A heating rate of  0.1 K d$^{-1}$ 
corresponds to $\sim 1.9\times 10^{-3}$ erg cm$^{-3}$ s$^{-1}$,
which is to be compared with the GCR rate of $\sim 1.7\times
10^{-7}$ erg cm$^{-3}$ s$^{-1}$.
(This assumes an ion pair production of 30 cm$^{-3}$ s$^{-1}$
at this altitude [from Nicolet 1975] and an energy expended
per ion pair of $\sim35$ eV [from Porter et al. 1976].)
 Our {\it  ad hoc }
     increase by 100 in the GCR rate for  $D_{\rm SN}=10$ pc
 raises this volume  heating rate 
     to a level that is still only
$\sim 0.01$ of the background rate, which is negligible.
%Even at these
%relatively small heating rates, the cosmic and gamma-rays
%would only increase the atmospheric heating about 1\%
%at $D_{\rm SN} = 10$ pc.  SNe
  A 
 SN  would have to be extremely close
(i.e., $\la 1$ pc)
    to
drastically perturb the atmosphere through heating.

The focus of this study was on the $NO_y$ production.  Both the
cosmic and gamma-rays should also produce $HO_x$ (e.g., see
Muller \& Crutzen 1993).  We tried a sensitivity study including
the $HO_x$ source in our cosmic ray simulation and assuming
2 $HO_x$ constituents are produced per ion pair.  This simulation
was only modestly different from the $NO_y$-only computation
and showed reductions of about 1\% or so in the calculated total
ozone depletion.  The reductions in the ozone depletion were
caused by interference from the $HO_x$ family constituents with 
the $NO_y$ family constituents through reactions such as

\hspace{0.5in}$OH$ + $NO_2$ + $M$ $\rightarrow$ $HNO_3$ + $M$. \hspace{0.5in}\hfil  (K)

Other potential uncertainties relating to the atmospheric physics
    include: 1) input boundary conditions for the
source gases, the progenitors of the more reactive atmospheric
constituents; 2) reaction rates and photodissociation cross
sections; 3) cosmic ray ionization rates from Nicolet (1975)
used for scaling the SN input.

% Concerning the astrophysics 
% ; 4) cosmic ray fluxes
% from SN; 5) gamma-ray fluxes from SN; and 6) duration
% of SN effects.

The astrophysics portion of the calculation involves additional uncertainties.
  For instance, the {\it ad hoc} factor of 100 increase in the
GCR rate for $D_{\rm SN}=10$ pc was chosen simply so as to produce a CR flux
   representative of that expected from a SN,
   and in fact the fluence
integrated over 20 yr amounts to $\sim 10^{50}$ erg.
   (As noted earlier, however,
the 20 yr run time in our model was chosen  to ensure steady state 
   would be achieved.)
   Obviously in a more realistic scenario,
     the increased flux of
cosmic rays at the top of the atmosphere would not lead to a uniform
  elevation of the ionization rate at all altitudes;  a proper
calculation would involve solving
      the  diffusion of the enhanced flux to lower altitudes.
  This is beyond our current level of sophistication.
 In effect, by uniformly augmenting the ionization for all altitudes,
      we are probably over-representing the increased
ionization at the lower altitudes.
   Second,  for  the gamma-ray flux  we have also avoided the diffusion
(i.e, radiative transfer) problem by adopting simple energy-dependent
attenuation coefficients. Although our estimate of the fractional
energy deposition versus altitude agrees quantitatively
    with results from a  proper
radiative transfer calculation, there may be other intricacies of the
process which are not captured in our technique. 
   Finally, there is considerable  uncertainty in the duration
 and functional form (in time) of
  both the   cosmic ray  and gamma-ray fluxes.
% For generality we have
% adopted a simple ``on/off'' prescription,  with ``on'' times
% of  20 yr for cosmic rays and 300 d for gamma-rays. 
 We find that the time 
             to reach steady state in both sets of  calculations
is relatively short, thus it appears that ``flux''  is more relevant than ``fluence''
  for our results.
    Values
for the cosmic ray acceleration time $\Delta t_{\rm acc}$
  within the 
SN remnant  ranging
from $\sim10$ to $\sim10^5$ yr have been quoted 
in the literature 
  (Blandford \& Eichler 1987, 
    Ellis \& Schramm 1995).
Our model run time of 20 yr
is toward the lower end of this range, therefore if we were to take the
additional step of assigning  a physical significance to our model run time, 
      our calculated
ozone depletion would represent in some sense an upper limit:
     larger values of
    $\Delta t_{\rm acc}$  would dilute
the cosmic ray energy in time to a greater extent, and lead to
  smaller ozone depletions.
  We feel confident in being able to scale
to other values of total adopted energy and ``on'' time
        based on the limited number
of runs which we calculated.  
    Given the current crude level of
understanding of cosmic and gamma-ray fluxes from SNe, it was thought
that more complicated assumptions involving additional free parameters,
such as power law or exponentially decaying forms for the fluxes, 
  would be unwarranted.

\section {Discussion }

%It is difficult to say with any degree of certainty
%how significant the ozone depletion must be in order to make
%a serious impact on life;
%  we take a factor of $\sim2$ increase
%in the transmitted ``biologically active''
%1 UV radiation to be an appropriate fiducial benchmark
%(see Table 1 and Fig. 2 of Madronich et al. 1998).

In order to quantify the impact on life of different
levels of ozone depletion,
we use the results of Madronich et al. (1998).
 From their Table 1 and Figure 2,
     we take a factor
of $\sim2$ increase in the biologically active UV flux
to be a threshold for significant biological effects.
This increase corresponds to an ozone depletion of $\sim47$\%,
given that the biologically active UV flux scales roughly as
the reciprocal of the ozone column (Madronich et al. 1998).\footnote{
Although the exponential, or Beer-Lambert model, for
attenuation is a good approximation at a given wavelength,
it is not appropriate for biological
effects at the Earth's surface. 
  The biological spectral sensitivity
  depends on the integrated UV
    flux between about 290 nm and 330 nm.
 One can show analytically 
that the biologically active UV flux varies roughly
as the reciprocal of the ozone column
 given that (i) the ozone absorption cross section decreases exponentially with wavelength
in this spectral region (290--330 nm), and
(ii)  the biological action spectrum decreases exponentially with wavelength over
this same spectral region  (Madronich et al. 1998).
}
There will probably be an interference between the increases
in $NO_y$ caused by the gamma-rays and cosmic rays, therefore
   a summation of the
separately computed ozone depletions from the two effects will result
in a maximum estimate of the total SN influence.
% Although there will probably be a minor interference
% between the increases in $NO_y$
% caused by the gamma-rays and cosmic rays, a summation
% of the separately computed ozone depletions from the two
% effects will result in a reasonable estimate of the 
% total SN influence.  
  Summing our gamma-ray and cosmic ray depletions
for  $D_{\rm SN} =10$ pc, and taking into account that our
adopted energy is larger than that found in the latest SN study
mentioned earlier, we obtain a fiducial ``critical distance'' 
to significantly disrupt ozone of $D_{\rm crit}\simeq 8$ pc
for a SN with a total gamma-ray energy $\sim1.8\times 10^{47}$ erg.

The impulsive addition of energy into the atmosphere
by other types of catastrophes may also lead to the
production of $NO_y$, therefore the accompanying
destruction of ozone may not be unique to the SN
scenario. A geometric dilution factor of 
$\sim10^{-22}$ for the cross section of the Earth
 as seen from
   10 pc means that the SN gamma-ray energy of
$2\times10^{47}$ erg translates to $\sim2\times 10^{25}$
erg intercepted by the Earth.
 An asteroid with mass $\sim3\times10^{17}$ g and
speed $\sim25$ km s$^{-1}$
such as the one responsible for the Cretaceous-Tertiary
extinction (Alvarez et al. 1980)
has $\sim10^{30}$ erg of kinetic energy (KE),
so that even if as little as $\sim10^{-5}$ of the KE
were channeled into atmospheric heating,
it would rival the SN input.
 On the other hand, the creation of  $NO_y$
through a sudden event would probably be less long lasting:
First, the transport of  $NO_y$
 from the stratosphere to the troposphere
would occur over a few years. Tropospheric
 $NO_y$
exists mainly as $HNO_3$,
and would be removed by precipitation.
Second, a localized event in the atmosphere
would probably not be communicated globally.
For instance, if an asteroid generated  $NO_y$
along its trajectory through the atmosphere,
it would not necessarily mix globally on a time
scale short compared to its destruction time scale.
A period of sustained energy input is required
to maintain
continuously high levels of  $NO_y$
if the depletion of ozone is to be effective.
   On the other hand, the asteroid scenario is known
     to
have many other disastrous  effects in terms of 
atmospheric pollutants. For instance the introduction
of massive amounts of sulfur, leading to sulfuric acid,
is thought to be devastating to life.

For a nearby SN to disrupt life on Earth,
there must be some finite probability
that such an event has occurred
within the past
several hundred   million years.
For our galaxy the SN rate is $\sim1.5$ per century
(Cappellaro et al. 1999).
Clark et al. (1977) estimate the Sun should pass
within 10 pc of a SN during each spiral arm passage,
implying a SN rate of $\sim10$ Gyr$^{-1}$.
Two key issues affecting the SN rate are
(i) the vertical scale height $h_{\rm SN}$ over which
supernovae (SNe) are distributed, and (ii)
the displacement of the solar system from the galactic
plane. The study of Clark et al., for example, ignores
the vertical disk structure and concludes that every passage
through a spiral arm results in a close-proximity
SN.
 An opposing view is presented by van den Bergh (1994)
who adopts $h_{\rm SN}=300$ pc, thereby giving
 $\sim0.3$ Gyr$^{-1}$ for SNe within 10 pc.
 Maiz-Apellaniz (2001) utilizes a sample of O-B5 stars,
precursors to core-collapse SNe, obtained from the 
{\it Hipparchos} catalog to determine $h_{\rm SN}=34$ pc.
He also finds the Sun to be $\sim24$ pc above the galactic
plane $-$
a value that is fortuitously less than $h_{\rm SN}$.
The value $h_{\rm SN}\simeq30$ pc
is a factor of $\sim10$ less than that used by
van den Bergh, and leads to a SN rate
intermediate between van den Bergh and Clark
et al., i.e., $\sim3$ Gyr$^{-1}$  for $D_{\rm SN}<10$ pc,
or $\sim1.5$ Gyr$^{-1}$  for $D_{\rm SN}<D_{\rm crit}\simeq 8$ pc,
where we used the fact that the rate varies as $D_{\rm SN}^3$.

If the cosmic ray flux scatters off magnetic field
  irregularities
in the interstellar medium, a spread in the arrival times
with increasing distance is introduced so that the
fluxes vary as  $D_{\rm SN}^{-4}$ rather than  $D_{\rm SN}^{-2}$.
This comes about because one has a factor  $D_{\rm SN}^{-2}$
due to the random walk in  spatial spreading, in addition to the
normal  $D_{\rm SN}^{-2}$ geometric factor  (Ellis \& Schramm 1995).
Benitez et al. (2002)
 argue that the Earth lies within a bubble of
hot gas caused by recent SNe which shields us from the
ISM magnetic field, and therefore leads to  $D_{\rm SN}^{-2}$ 
cosmic ray SN fluences.
%
%
%  One potential caveat to this line of reasoning
%in which the assumed SN fluences vary as  $D_{\rm SN}^{-2}$
%is  the possibility that the Earth may lie within a bubble of
%hot gas caused by recent SNe, so that when a new SN occurs
%the cosmic ray influence within the bubble may be more
%isotropic and therefore stronger.
%
     Benitez  et al.
 make this argument concerning
recent SNe in the Local Bubble and the Scorpius-Centaurus
OB association which may have generated $\sim20$ SNe
in the last $\sim10$ Myr.
  The cosmic ray flux $\phi_{\rm CR} \la 1.4\times 10^7$
erg cm$^{-2}$ y$^{-1}$   they quote
 using the minimal distance for the Sco-Cen SNe
$D_{\rm SN} \ga 40$ pc (and assuming a typical cosmic ray
acceleration time $\sim 10$ yr)
    is actually about a factor
of 10 less than we adopt in this work for $D_{\rm SN} = 10$ pc,
  leading us to believe the effect discussed
by Benitez et al. may produce ozone depletions
of only a few percent.
  The complications introduced by
having the Earth be inside a hot bubble
 which may enhance the cosmic ray flux
    during an interval
in which several successive SNe occur are interesting
and worthy of more detailed studies.
It may be the case, for example, that the cumulative effect
of having many SNe occur  within a relatively short interval
     is to elevate the  general cosmic ray flux
within the Local Bubble to a level higher than simple
      $D_{\rm SN}^{-2}$
fall-off from a single SN 
would suggest.

\section {Conclusion }

In summary, we have calculated detailed
atmospheric models to determine the extent
of the reduction in ozone due to elevated levels
of odd nitrogen  induced from gamma-rays
and cosmic rays produced in a local SN.
% Our study represents an advance over
% earlier  work:
   Our procedure is as follows: 
(i) We  utilize a detailed 2D model
for the Earth's atmosphere,
incorporating the latest advances in photochemistry
and transport.
(ii) For the gamma-ray spectrum we take as input the
observed spectrum from SN1987A, scaling the total energy
to the most recent value determined by workers investigating
core-collapse SNe with a red supergiant progenitor.
(iii) For the cosmic ray spectrum we adopt scaled-up values
of the empirically observed ionization rates
in the atmosphere from galactic cosmic rays.
(iv) To estimate the frequency of local SNe we utilize
recent estimates of global SN rates for spiral galaxies
and the results of a recent investigation into
the vertical spatial extent of core-collapse SN progenitors.
   Our primary finding is that a core-collapse SN would need
to be situated approximately 8 pc
away to
produce a combined ozone depletion from both gamma-rays
and cosmic rays of $\sim47$\%,
which would roughly double the globally-averaged,
biologically active UV reaching the ground.
The rate of core-collapse SNe occurring within 8 pc
is $\sim1.5$ Gyr$^{-1}$.
As noted earlier,  our calculated ozone depletion
% for $D_{\rm SN}=10$  pc
 is significantly  less than that found by Ruderman (1974),
and consistent with Whitten et al. (1976).
Given the $\sim0.5$ Gyr time scale for multicellular
life on Earth, this extinction mechanism appears to be 
less important than previously thought.

It is a pleasure to acknowledge
stimulating conversations with Frank Asaro,
Narciso Benitez, Adam Burrows, Enrico Cappellaro,
Aimee Hungerford, Sasha Madronich, Frank McDonald,
Joan Rosenfield,
John Scalo, David Smith, Floyd Stecker, Sidney van den Bergh,
William Webber, Craig Wheeler, and  Stan Woosley.

\vfil\eject
\centerline{ FIGURE CAPTIONS }
%\medskip

Figure 1. 
 Contour plot of global average of ozone depletion 
during Year 2 from gamma irradiation, for $D_{\rm SN} = 10$ pc
and $i_{\rm SN} = 0^\circ$.

Figure 2.
  Contour plot of global average of ozone depletion during Year 20
due to the cosmic irradiation, assuming an enhancement of 100-fold
in the empirically observed GCR  ionization rate, and $i_{\rm SN} = 0^\circ$.

Figure 3. 
   Peak annual average reduction in global ozone from gamma
irradiation for $D_{\rm SN} = 10$ pc as a function of $i_{\rm SN}$
(average over Year 2).

Figure 4. Annual average reduction in global ozone from gamma
   irradiation from a SN as a function of  $D_{\rm SN}$,
assuming  $i_{\rm SN} = 0^\circ$ (average over Year 2).
 The solid line indicates a dependency of  $D_{\rm SN}^{-2}$.

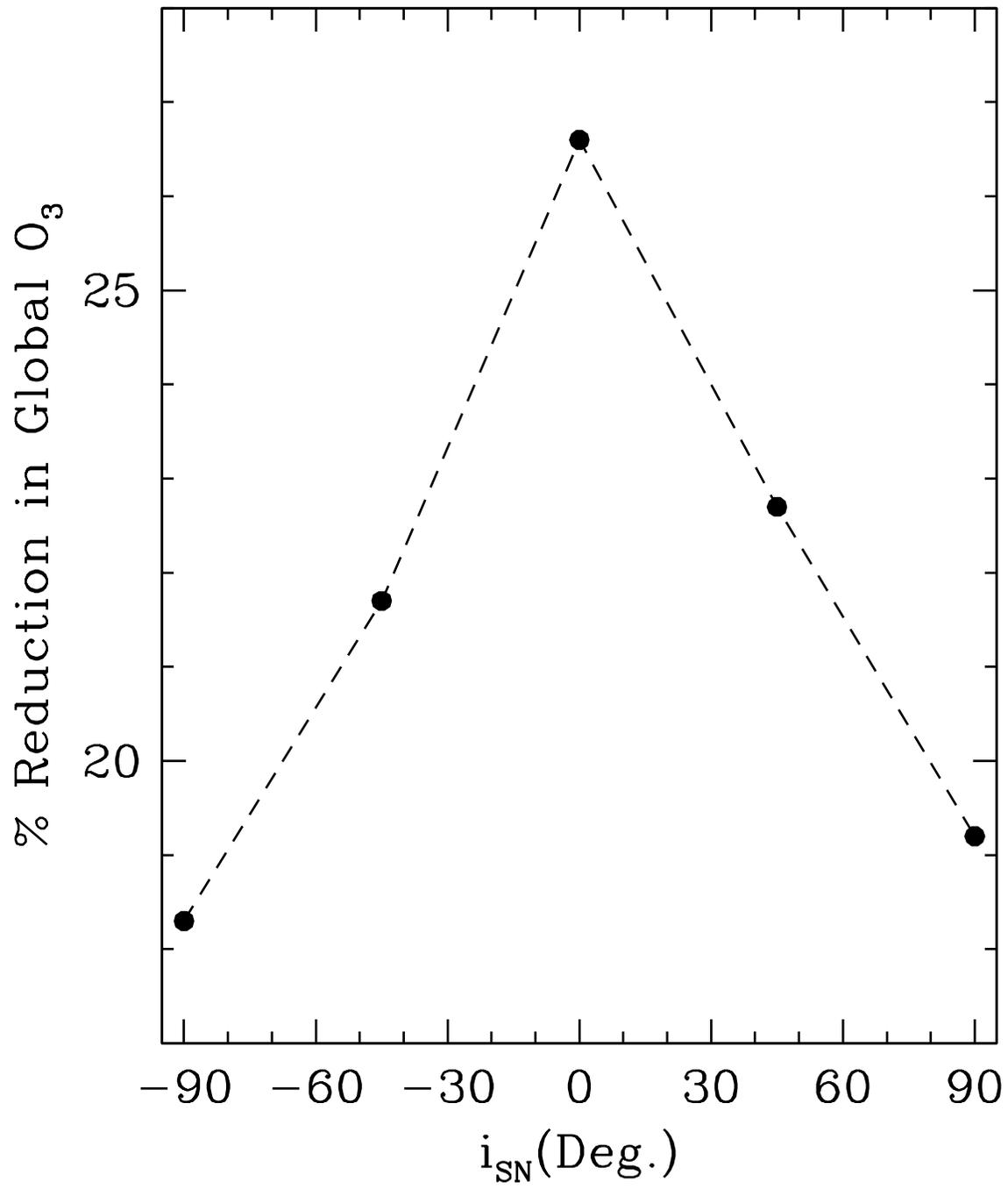
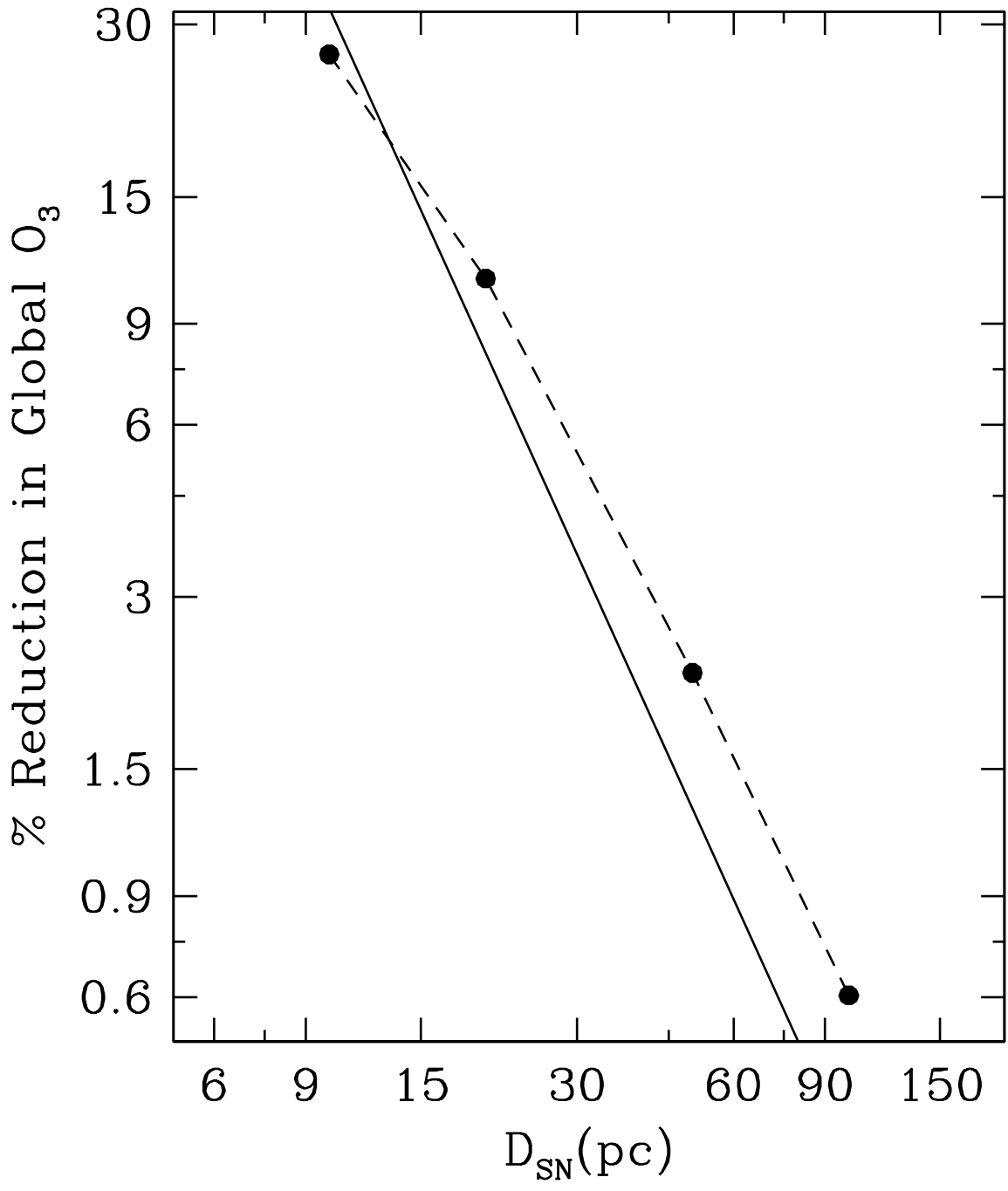
\end{document}